*Chapter 10*

# Demand for shared mobility to complement public transportation: Human driven and autonomous vehicles

*Shadi Djavadian[1] Bilal Farooq[2] and Seyed Mehdi Meshkani[2]*

*In:*

## Shared Mobility and Automated Vehicles: Responding to socio-technical changes and pandemics

*Edited by: Ata Khan and Susan Shaheen*

Recent advances in communication technologies and automated vehicles have opened doors for alternative mobility systems (taxis, carpool, demand-responsive services, peer-to-peer ridesharing, and car sharing, shared autonomous vehicles/shuttles). These new mobility services have gathered interest from researchers, public and private sectors as potential solutions to address last-mile problem—especially in low density areas where implementation of high frequency buses is not feasible. In this study we investigate the effects of ride-sharing service on travel demand and welfare, as it complements public transportation under different scenarios. Two types of management and vehicle types are considered: crowdsourced human driven vehicles (HDV) (e.g. Uber, Lyft) and centrally operated shared autonomous vehicles (SAV). The influence of fare discount on demand and mode shift is also investigated. A case study of Oakville road network in Ontario, Canada is conducted using real data. The results reveal that ride-sharing having the potential of increasing ridership by 76 per cent and decreasing wait time by 47 per cent under centrally operated shared autonomous vehicles with 50 per cent fare discount for sharing.

---

[1] Ford Mobility
[2] Laboratory of Innovations in Transportation (LiTrans), Department of Civil Engineering, Ryerson University

## 10.1    Introduction

Among the societal impacts caused by increasing urbanization over the coming decades, the transportation sector acts both as a driver and a recipient of change: better access to mobility makes cities attractive, while ever growing demand will cause capacity issues that even massive investments into new infrastructure cannot cope with. Already today, congestion in and around urban areas in the EU costs nearly 100 billion Euros, or one per cent of the GDP, per year [1]. On the other hand, in rural areas, serving the transportation needs of a shrinking and aging population efficiently becomes increasingly challenging [2]. Likewise the forecasted cost of congestion to economic in the form of GDP for the Greater Toronto and Area which is already the 6th most congested city in the world [3] will balloon to $7.2 billion by year 2031 [4].

One solution would be to provide high frequency and more accessible public transit systems to mitigate congestion and remove private cars from the roads. However, the demand for transit, especially in low density areas, is often hindered by absence of well-organized and effective door-to-transit station service known as the "first mile/last mile" problem [5]. New technologies, in particular shared on-demand flexible transit and shared autonomous vehicles (SAVs) (e.g: Optimus Ride, drive.ai, Waymo, EasyMile, Bestmile, etc.) [6], create an opportunity to break this vicious circle by offering new and fundamentally more flexible mobility services as a compliment to fixed-route transit [7].The effectiveness of shared autonomous vehicles compared to privately own vehicles have been studied using simulation for both European cities (e.g: Berlin and Lisbon) and American cities (e.g.: Austin). The results showed that one SAV has the potential of replacing the demand served by ten privately owned conventional vehicles [8, 9, 10].

Already with conventional human-driven vehicles, transportation network companies (TNCs) have started to disrupt the mobility sector and are gaining market share at the expense of traditional taxi services and public transit. Many public transport operators, either alone or in public-private partnerships, are already exploring ways to provide new mobility services and are piloting different types of so-called flexible transportation systems, including demand-responsive shuttle, ride pooling, or micro-transit services or shared autonomous shuttles. For the list of current public and private partnership projects on shared autonomous vehicles in US the interested reader is referred to APTA [11]. A recent example of such services in Ontario is City of Belleville, Ontario, that since September 2018 has partnered with the Toronto-Based company Pantonium [12], in which city's bus service acts more like a ride-hailing service from 9pm-12am, during weeknights. The reason behind this experiment is that city of Belleville has a population of 50,000 where annual transit ridership is equal to daily transit ridership in Toronto, as such providing a high frequency fixed route traditional transit service is not a feasible solution. However, flexible transit system can be a viable solution for a low dense city such as Belleville. Due to the success of initial pilot project, the city is now considering changing more routes to flexible service as well as decreasing the reliance on 40-foot diesel buses driving around on fixed routes [13]

However, one of the main obstacles for the introduction of a new type of transportation service is the lack of adequate demand models. Having a reliable estimate of demand is necessary to scope, design, and plan all aspects of the service, ranging from fleet sizing, determining hours of operations, level of service, fleet dispatch strategies, and pricing. Getting all of this right from the beginning is crucial to ensure a successful introduction and uptake, as well as the continued economic viability of the service.

Majority of studies on shared mobility (e.g. ride-sharing, car sharing) have either considered demand fixed as exogenous input trying to optimize profit, fleet size, station location and reservation policies of shared mobility [14], or considered fleet size fixed and as exogenous input trying to model interaction of demand with supply [15] . However, as shown by Djavadian & Chow [16], FTS behaves as a two-sided market and failure to consider the adjustment process of both the operator and travelers and their interaction leads to over estimation of demand, their impacted welfare and fleet size.

Djavadian and Chow [16, 17], proposed a simulation-based method to evaluate the market equilibrium for FTS in one and two-sided market by adopting an agent-based day-to-day adjustment process designed to reach an agent-based stochastic user equilibrium (SUE). However, in their studies, they only evaluated the dynamic FTS for a HDV single ride service. In this study we extend the work of Djavadian & Chow [16] to shared HDV and AV services. For comparison purposes a case study similar to [16] is conducted using Oakville road network.  The aim of this study is to evaluate the demand and social welfare effects of shared HDV and AV services under different dynamic operating policies (e.g. fleet vehicle capacity, fare price, minimum wage of drivers).

The remainder of this paper is organized as follows. In the background section an overview of agent-based day to day adjustment process for evaluating dynamic FTS will be presented. The methodology section presents design of our case study to evaluate dynamic shared HD and AV. After results and analysis are presented followed by summary and future work directions.

## 10.2    Background

Public policy makers are often faced with questions such as: what fleet size should be employed and what dispatch algorithm should be used for the given fleet size? Should they offer single ride or shared ride? Should such a service be operated 24 hours a day, or only during peak period? What pricing scheme should be used? Each of these design decisions and operating policies leads to different level of service (LOS) which in return might lead to different demand and different impacted welfare.

In literature, flexible transit specifically taxi has been extensively studied. The focuses of these studies have been on fleet size, pricing and vehicle routing problems. These studies mostly look at taxi service or other flexible services from the point of view of service operators where the aim is to minimize operators' and travellers' cost as such only the within day dynamics are considered and it is assumed that the system is at steady-state, meaning that current condition will be the same tomorrow and travelers will make the same choices the next day. The supply-demand equilibrium considered in these studies is based on market equilibrium where the learning behaviour of travelers is not considered. However as mentioned by Quadrifoglio & Li [18] the demand for flexible transit varies according to the level of service of flexible transit. It may also change from one day to another and this is a key factor when it comes to transportation planning and is of great importance to public agencies when trying to justify one alternative over another.

Flexible transit system as noted by Djavadian & Chow [16, 17], is fundamentally different from general transportation systems: a) the system performance is dependent on both the choices of the travelers as well as the policies adapted by the FTS serving as an additional decision-maker, b) unlike traffic network where the route choice exclusively depends on the travelers in the case of FTS passenger's route is decided by the operating policy of the FTS. In return the operating policy of the FTS is also affected by choice sets of all travelers. c) as opposed to traffic network cost function which is monotonic, it has been shown in the literature (e.g. [19]) that demand responsive public transit cost function can be non-monotonic with respect to flow. The link costs in an FTS are dependent on the operating policy and may be non-monotonic or follow discrete step functions. Due to inherent characteristics of FTS, a framework to evaluate different designs of FTS needs to meet the following criteria: a) Capture heterogeneity of travelers, b) Capture interaction between travelers, c) Capture impact of operation policy on travellers' choice, d) Capture impact of travellers' choice on FTS level of service and e) Capture day-to-day learning process of both <u>travellers</u> and FTS <u>drivers</u> (two-sided market). Because of the complicated dependencies posed by the FTS as defined, a steady state model would not be able to model the sensitivities attributed to within-day dynamic operating policies as desired. Therefore, day-to-day model needs to be used.

In their study, Djavadian & Chow [16, 17], proposed an agent-based day-to-day adjustment process for evaluating the effects of design parameters and operating policy of flexible transportation system on equilibrium demand and their impacted welfare. They demonstrated that the interaction between the operator and travelers was equivalent to a two-sided market made popular by companies like Airbnb and Uber, where the trading platform is the spatial network. This was done so that the operating decisions of the service operators become part of the output of the modeling framework. The proposed process incrementally adjusts demand levels interacting with a stochastic dynamic system operator in search of an invariant distribution. Case studies using real data from Oakville, Ontario, as a first/last mile problem example demonstrated the sensitivity of the one sided and two-sided day-to-day models to operating policies.

The approach proposed by Djavadian & Chow [16, 17] is organic and supportive of the advanced mobility systems that are taking place in our society due to rapid urbanization and population growth. However in their study, the focus was only on single-ride FTS driven by human drivers. In this study the framework is extended to FTS service under HD and AV setting, providing ride-sharing service as opposed to single-ride service.

## 10.3    Methodology

To answer the research question posed in the Introduction section, we developed and implemented ride-sharing and autonomous vehicles in the original framework (see **Figure 10.1**).

**"Figure 10.1 ABOUT HERE".**
*Figure 10.1. Agent-based transportation simulation tool framework*

## 10.3.1 Ride-sharing

Flexible on-demand ridesharing refers to a mode of transportation which allows users to share a vehicle and car related expenses such as gas, toll, and parking fees with others that have similar itineraries and time schedules [20]. In fact, ridesharing is a system that can combine the advantages of private cars and fixed-line systems; that is, flexibility and speed of private cars with the reduced cost of fixed-line systems, at the expense of convenience ( [20]). Saving travel cost, reducing travel time, mitigating traffic congestions, conserving fuel, and reducing air pollution are a number of ridesharing benefits for both participants (drivers and passengers) and society as well as the environment [20]. Examples of flexible and shared use mobility are systems like demand-responsive transit, Uber, and Zipcar, where the performance of the system is state-dependent and stochastic.

**10.3.1.1 Dynamic Dial a Ride Problem**

Since 1970 various vehicle routing policies have been studied for a dial-a--ride-problem (DARP), focusing on both static and dynamic problems. What distinguishes, dynamic dial-a-ride problem from the static dial-a-ride problem is that in dynamic dial-a-ride problem vehicles' routes are modified in real-time in response to trip requests arriving in time. The dynamic dial-a-ride problem usually has two conflicting objectives identified as (a) system efforts and (b) the customer's interests [21].

In this study the dynamic DARP proposed by Hyytiä et al [21] is adapted. The same model was used by Djavadian & Chow [16], however in their study they set the capacity of fleet size to 1 and considered only single ride FTS. Our microsimulation implementation involves fleet vehicles updating their path en-route to accommodate added pick up and drop offs dynamically.

At the beginning of each day a fixed set of $v$ uncapacitated vehicles with constant speed is assumed to provide pickup and delivery service for customers. The total fleet size on each day $\Lambda_d$, depends on the policy of service provider. In this study two types of fleet vehicles and management styles our considered, centrally operated AV fleets and crowdsourced HDVs. The strategy for updating $\Lambda_d$ for AVs is explained in the next section, whereas the strategy for updating for the latter is taken from Djavadian & Chow [16].

Within day vehicle to customer assignment is as follows. When a trip request is placed, customer is assigned to a specific vehicle immediately, and the vehicle's route plan (list of nodes/intersection to visit) is then updated to include both the pickup location and delivery location in that order, with no request ever being rejected. An example of a tour for a ride-sharing problem is: $\xi = \{1\ 2\ -2\ 3\ -1\ -3\ 0\}$. The numbers represent customers' ID in the request table list, where as positive value denotes pickups, negative values denote drop offs, and 0 denotes depot. An example for single ride service tour is: $\xi = \{1\ -1\ 2\ -2\ 0\}$. Since this service is shared-use as a dynamic DARP, a customer may be delayed in being dropped off in favor of another customer if it minimizes total cost.

As mentioned before, $\xi$ for each active vehilce $v \in \Lambda_d$ is updated using the model proposed by Hyytiä et al [21] where they considered the non-myopic dynamic DARP as a multi-server queue problem. The model is being non-myopic since decision is not only based on current information, but it also takes into account the future conditions with steady state queue characteristics. The model allows user to choose between myopic and non-myopic assignment. Since last mile/first mile problem is considered in this study and travelers are assumed to have defined departure time as opposed to random, in this study myopic assignment is used. $\kappa = 0$ refers to a myopic system whereas $\kappa > 0$ refers to a non-myopic system.

In their study, Hyytiä et al [21]assumed the process of modeling vehicle assignment to be similar to assigning a customer to a server in a multi-server queue and they developed a policy called mm1 that aims to minimize a weighted sum of the mean passengers' travel time and distance the vehicles travel (per passenger) as shown in **Equation 10.1.**

$$mm1: argmin_{v,\xi}[C(v,\xi) - C(v,\xi')]  \qquad (10.1)$$

**Equation 10.2** presents the calculation for relative cost of vehicle-route pair $(v, \xi)$ as a sum:

$$C(v,\xi) = \gamma T(v,\xi) + (1 - \gamma)\left( \kappa T(v,\xi)^2 + \sum S_i(v,\xi) \right) \qquad (10.2)$$

Where:
$\xi$ is new tour and $\xi'$ is prior tour (the tours are constructed using a TSPPD heuristic, C : cost value
$T(v,\xi)$: the current work backlog of vehicle $v$ (measured in time), and
$S_c(v,\xi)$: denotes the residual sojourn time of a customer $c$ in the system.
$\gamma$ : corresponds to combination of the minimization of the system's effort and travelers costs and can take values between 0 and 1. In this study value of 0.5 is used, which takes into account both system's effort and the travel time cost incurred by the traveler.
It is worth mention that the DARP implemented in this study based on Hyytiä et al [21] is with centralized dispatch.

**10.3.1.2 Fleet vehicle path updating**

In the study conducted by Djavadian & Chow [16, 17] since the purpose was to model single ride, the microsimulation only updated path of fleet vehicles only at pick up and drop off locations and the path remained the same in between. However, to accommodate ride-sharing the following changes are made to the microsimulation path finding:

For example, let's assume there are two commuters, 1 and 2. With 1 having origin/destination pair (A-C) and 2 having origin/destination pair (D-C) (**Figure 10.2**). The request time of 1 is at 8:00am. Further let's assume fleet size of 1 with capacity 1. When commuter 1 makes a request the tour of vehicle is {1 -1 0} and the path is {Depot, A, B, C, E, Depot}. However, let us assume that commuter 2 makes a request when vehicle carrying commuter1 is on link connecting A & B. In the case of single ride [16, 17] , the tour and path of the fleet vehicle would have been updated as follows: {1 -1 2 -2 0}, {Depot, A, B, C, E, B, C, E, Depot}. In the case of single ride, the entire path from start to end is calculated. However, in the case of shared-ride only the next intersection on the list is known to the vehicle since it is possible for the next node to change. For example, at 8:00am the tour of vehicle is {1 -1 0}, therefore the path is {Depot, A}. Once the vehicle arrives at node A, it updates the next node on its path to destination C which is B (based on shortest travel time) so the new path becomes {Depot, A, B}. Once travelling on the link between A and B, it is notified of new passenger pick up at D, so the tour is changed to {1 2 -1 -2}. As such when vehicle arrives at B, it looks for shortest path to D and the next node on its path which is E, as such the path becomes {Depot, A, B, E} and so on. This not only allows us to implement ride-sharing, but it also allows us to further extend the framework to dynamic distributed routing, similar to Djavadian & Farooq [22] where AVs are guided by a network of intelligent intersections.

**"Figure 10.2 ABOUT HERE".**

*Figure 10.2 Sample network*

## 10.3.2  *Shared Automated Vehicles*

AV has the power to dramatically change transportation systems operations [22]. Increased safety, convenience, productivity, traffic efficiency, lower congestion, and reduced environmental impact can be considered as the main benefits of AVs [7]. AVs also have another potential benefit: they have the ability to further facilitate car-sharing and ride-sharing behavior, since they can overcome some key barriers—especially the limited accessibility and reliability of today's car-sharing and ride-sharing programs [7]. Shared autonomous vehicles enable people to call up distant SAVs using mobile phone applications. Moreover, they can anticipate future demand and relocate in advance to better match vehicle supply and travel demand. [7].

An on demand flexible transit system can be operated by a decentralized operator or a centralized operator. For the purpose of this study it is assumed that the centralized operator is a public agency with maximum *M* registered automated vehicles. Unlike two-sided FTS with human driven vehicles (HDVs) [16], in the case of FTS with automated drivers, the central operator is the main decision maker and decides how many vehicles should be used on each day (**Figure 10.3)**. The green square in Figure 10.3 highlights the difference between crowdsourced HDV (dotted red square) and centrally operated AV in this study. What separates centrally operated SAV in this study from the centrally operated HDV in Djavadian & Chow [17] (see Figure 1a) is that in their study the day-to-day adjustment process of only travelers was considered. The fleet size remained the same from one day to another, whereas in this study to incorporate more realistic behaviour, fleet size is changed from one day to another depending on demand level and feedback from the system. As a public service provider, the aim is to serve the citizens, however maintaining AVs is also costly, and as such the strategy of central operator on each day *d* is to reach social optimality where the needs of both parties are met. The results from Djavadian &

Chow [16] showed that for each fare price, social optimality occurs when there is a perfect match between supply and demand (market equilibrium). Therefore, to reach system optimality, **Equation 10.3** is used to adjust fleet size of AVs from one day to another.

$$\Lambda_d = \frac{\Upsilon_d}{\varpi}, \ 0 \leq \Lambda_d \leq M \tag{10.3}$$

where:

$\Upsilon_d$: Predicted total taxi demand on day *d.* For simplicity in this study it is assumed that $\Upsilon_d = \Theta_{d-1}$, where $\Theta_{d-1}$ is actual taxi demand on day *d-1*. This means that the operator assumes that the demand on day *d* is the same as demand at day *d-1*. This is a myopic estimation of demand, however as shown by Djavadian & Chow [16, 17], the day to day adjustment process for evaluating FTS eventually reaches stochastic user equilibrium where demand for day *d* is equal to demand for *d-1*.

$\varpi$: capacity of each AV vehicle, in this study capacity of both HDV and AV is set to 4-passenger max. We also investigated the 7-passenger vehicles but found no major differences.

**"Figure 10.3 ABOUT HERE".**

*Figure 10. 3 Agent-based day-to-day SUE framework for centrally operated AV fleets*

## 10.4    Case Study

The modified agent-based day-to-day adjustment process is applied to a hypothetical ride-sharing flexible transportation system using real data obtained for Oakville, Ontario (**Figure 10.4**). The proposed ride-sharing FTS is used as a potential feeder service solution for the last mile problem connecting residents from home to the terminal rail station. The network chosen for this case study has been previously used for last mile study site by Djavadian & Chow [16, 17] and Alshalalfah & Shalaby [23], this will allow us to draw comparisons with the previous studies.

**"Figure 2.4 ABOUT HERE".**

*Figure 10.4 Oakville Network*

### 10.4.1  Case Study Objectives

Similar to the study conducted by Djavadian & Chow [16, 17], a taxi service is treated as a ride sourcing service. In addition, the focus of this study is also on residents of the town of Oakville who commute to downtown Toronto for work during the morning peak period by taking Go transit out of Oakville Station. However, there are four major differences between the case study presented in this study and the ones conducted by Djavadian & Chow [16, 17]:

   a) The dynamic with-in day routing policy used by Djavadian & Chow [16] only

focused on single ride service, whereas in this study ride-sharing option is introduced

b) In their study the focus was only on human driven FTS, whereas in this study we introduce shared-automated vehicles

c) The fleet size in Djavadian & Chow [16] depended on day-to-day strategy of the drivers themselves (entering the market or not) without any decision made by the operator of the FTS. However, in this study for the case of AVs the operator will decide to change the fleet size from one day to another

d) we investigate the effect of vehicle capacity on ride-sharing demand.

Since the focus of this study is on home-work trips, it is assumed that all the commuters are willing to share a ride with strangers. This is a reasonable assumption, as studies have shown that when it comes to work trips, commuters put more emphasis on travel time as opposed to with whom they share a ride [24] .

## 10.4.2  Problem Definition

For comparison purposes, similar to Djavadian & Chow  [16, 17], the focus of this case study is also on home to work (H-W) trips and considers five access modes to the Oakville station: bus, automobile, walk, fixed route transit and taxi (the modes listed in the Transportation Tomorrow Survey (TTS) [25]. During the study period (6:30-7:30), 2000 commuters access Oakville Go Station for work trips from Oakville to Toronto. Based on the statistics obtained from 2011 household survey by TTS [25] market share for different modes used by commuters to access Oakville Go Transit are as follows: 73 per cent used auto as the access mode, 19 per cent used bus, ~1 per cent used taxi (17 commuters), 6 per cent used bike and 1 per cent walked to Oakville Go Station. As can be seen from the access mode statistics given in the previous section, auto is considered a major access mode to Oakville Go Station and because of this high dependency on auto as an access mode to the station, a significant problem facing Go Transit in Oakville which is almost all its parking lots have reached capacity

For the purpose of this study, let's "assume" a transit public agency would like to provide commuters better access to Oakville Go Station (last/first mile problem). One way for the public agency to achieve this goal is to provide door-to-Go Station flexible transit service as studied by [23] and Djavadian & Chow ( [16], [17]). For the purpose of this study we assume that the "taxi" service serving the commuters is the public flexible transit service providing last/first mile service and that the public agency would like to improve the current available flexible transit service in the following ways:

- Improve level of service of current FTS by changing its design and operating policies:
  - Offering last mile/first mile ride-sharing service
  - Offering cost sharing to reduce fare cost
  - Using shared automated vehicles (SAVs) vs. crowdsourced ride-sharing service using human driven vehicles (HDVs)

The following questions for this case study will be answered:
  1. Effects of ride-sharing on demand and consumer surplus.

2.  Effects of fare discount on demand for ride-sharing
3.  Effects of automation on demand for ride-sharing

As mentioned previously for comparison purposes same network, base case population and calibrated Logit Model from Djavadian & Chow [16, 17] are used. Similarly, unless it is stated same values are used for agent-based day-to-day adjustment model parameters. A maximum fleet size of 10 vehicles is assumed for illustrative purposes. **Table 10.1** presents test scenarios for this case study. In total there are 8 scenarios, investigating effect of automation, fleet vehicle capacity and discount on equilibrium demand and welfare. Base case scenario (Day 0) is obtained from Djavadian & Chow [16].

*Table 10.1 Test scenarios attribute summary*

| Scenario | Fleet Type | Max Available Fleet Size | Fleet Size Capacity | Profit ($) Threshold | Fixed Fare Price | Fare Price ($)/additional 130m | % Fare Discount | Operating Cost ($)/Km |
|---|---|---|---|---|---|---|---|---|
| Base Case | HDV | 10 | 1 | 1 | 4.25 | 0.25 | 0 | - |
| HDV_0%* | HDV | 10 | 4 | 25 | 4.25 | 0.25 | 0 | 0.51 |
| HDV_15% | HDV | 10 | 4 | 25 | 4.25 | 0.25 | 15 | 0.51 |
| HDV_25% | HDV | 10 | 4 | 25 | 4.25 | 0.25 | 25 | 0.51 |
| HDV_50% | HDV | 10 | 4 | 25 | 4.25 | 0.25 | 50 | 0.51 |
| AV_0% | AV | 10 | 4 | - | 4.25 | 0.25 | 0 | 0.51 |
| AV_15% | AV | 10 | 4 | - | 4.25 | 0.25 | 15 | 0.51 |
| AV_25%** | AV | 10 | 4 | - | 4.25 | 0.25 | 25 | 0.51 |
| AV_50% | AV | 10 | 4 | - | 4.25 | 0.25 | 50 | 0.51 |

*HDV_0%: human driven fleet- 0 per cent discount for sharing the ride
** AV_25%: Automated vehicle fleet – 25 per cent discount for sharing the ride

## 10.4.3 Results

To illustrate the sensitivity of flexible transit demand to vehicle capacity, fare discount and operation management (crowdsourced HDVs or centrally controlled AVs), total of eight scenarios (**Table 10.1**) are considered with ten runs (days) for each scenario. The equilibrium state (demand, mode choice, departure time choice, desired arrival) from Djavadian & Chow [16] is used as base case scenario and starting point (Day 0) for each test scenario.

**Figure 10.5** presents shared FTS ridership demand over the range of 10 days for different discount levels and management/vehicle type (crowdsourced HDV and centrally operated AV). The results illustrate the effect of discount rate and management/vehicle type on ridership demand adjustment from one day to another. The first significant finding from the results as shown in **Figure 10.5** is the increase in ridership increase due to increase in vehicle capacity. In the base case scenario (single ride) the equilibrium ridership was 17. However, by increasing capacity to four and offering ride-sharing the ridership increased 76 per cent to as high as 30 from original 17. The notable shift in demand can be attributed to substantial decrease in wait time (47 per cent) and discount compared to base case as shown in **Figure 10.6**. Since this is a last mile/first mile problem, riders share either origin or destination (Go station) as such with a given fleet size it is more efficient to share a ride.

**"Figure 10.5 ABOUT HERE".**

*Figure 10.5 Ride-sharing demand*

**"Figure 10.6 ABOUT HERE".**

*Figure 10.6 Average wait time (min)*

The second noteworthy finding from **Figure 10.5** is that centrally managed AVs resulted in slightly higher demand compared to crowdsourced HDVs. The reason for this occurrence is because as shown by **Figure 10.7** the fleet size is more stable under centrally operated AVs as opposed to crowdsourced HDVs. The fleet size stability can be contributed to one main factor, which is updating strategy of the service provider. In the case of crowdsourced HDVs, the fleet size is determined by the number of drivers entering the market on each day, which depends on perceived profit of the drivers. If the perceived profit is high, more drivers will enter the market and vice versa. This fluctuation in fleet size will in turn results in fluctuations in demand, which will result in fluctuation in profit, which in long run converge to a stable point. Comparing **Figure 10.5** and **Figure 10.7** it can be seen that there is more obvious change in fleet size compared to change in demand, mostly in case of HDVs. This shows the sensitivity of HDV fleet to minor changes in profit.  However, in the case of centrally operated AVs by a public agency as we discussed before the aim is to reach social optimality, the fleet size is updated such a way that supply matches demand. Since central operator knows demand on each day it has the advantage of adjusting fleet size accordingly for the next day. This stability in fleet size in return results in stability in demand.

**"Figure 10.7 ABOUT HERE".**

*Figure 10.7 Fleet size*

From **Figure 10.5** it can be seen that introduction of discount into fare price has resulted in increase in ridership which is an intuitive. However, this increase in more pronounced under centrally operated AV fleets. The reason for this being that in the case of HDV fleet, higher discount means less profit for drivers as such number of drivers entering the market drops resulting in lower fleet size **(Figure 10.7)** or no drivers at all (case of 50 per cent discount). Conversely, in the case of centrally managed AVs, since the goal is to reach social optimal (match supply to demand) intendent of profit, the fleet size is not negatively affected by increase in discount rate.

**Figure 10.8** presents total consumer surplus under the 8 scenarios test. It can be seen that with the exception of HDV fleet size and 50 per cent discount, overall introduction of ride-sharing resulted in increase in consumer surplus with most increase being 2 per cent under centrally operated AV and 50 per cent discount. Under AV operated fleet and 50 percent discount, seven commuters switched from auto to ride-sharing and five commuters switched from regular transit to ridesharing. The results obtained shows the significant difference between two fleet size updating objective one being profit maximizing and one being social optimality.

**"Figure 10.8 ABOUT HERE".**

*Figure 10.8 Total consumer surplus*

**Figure 10.9** illustrates the difference between owning and operating AV fleet centrally and outsourcing rides in terms of profit. As can be seen from **Figure 10.9** the operator can accumulate more profit if its own and operates fleet itself. In the case of outsourcing the operator will only accumulate commission from the rides.

**"Figure 10.9 ABOUT HERE".**

*Figure 10.9 Total Profit*

## 10.5 Conclusion

In this study the agent-based day-to-day adjustment framework for evaluating dynamic FTS proposed by Djavadian & Chow [16, 17] is extended to include new and upcoming disruptive mobility services such as shared mobility and shared autonomous vehicles. A case study of Oakville, Ontario is conducted comparing the effect of ride-sharing on demand and consumer surplus with the single ride service. In total eight scenarios are tested with the aim of investigating the effect of fleet vehicle capacity, fare discount and management/ vehicle type (crowdsourced HDV fleet vs. centrally operated AV fleet) on travel demand and welfare. The results obtained highlight the presence of two-sided market where the choices of travelers are highly impacted by the choices made by the operators (level of service) and vice versa. One of the main finding is that introduction of ride-sharing with discount for the last mile/first mile problem resulted in 76per cent increase in ridership and 47per cent decrease in wait time and slight increase in consumer surplus. The comparison of centrally operated AV fleet with crowdsourced HDV fleet shows the important differences between profit maximizing service and social optimal seeking service and the effect on fleet size, demand and their impacted welfare. For the Oakville case study, under centrally managed AV fleet, the optimal case is when 50per cent discount is offered for sharing, which resulted in increase in demand, decrease in wait time, and increase in operator profit. Contrariwise, under crowdsourced HDV fleet the optimal case is when 25 per cent discount is offered. The results show that fleet size is more stable under centrally operated AV. In future will explore the role of differentiated services on the travel demand and welfare. We will also explore the performance of proposed framework on congested networks, e.g. downtown Toronto.

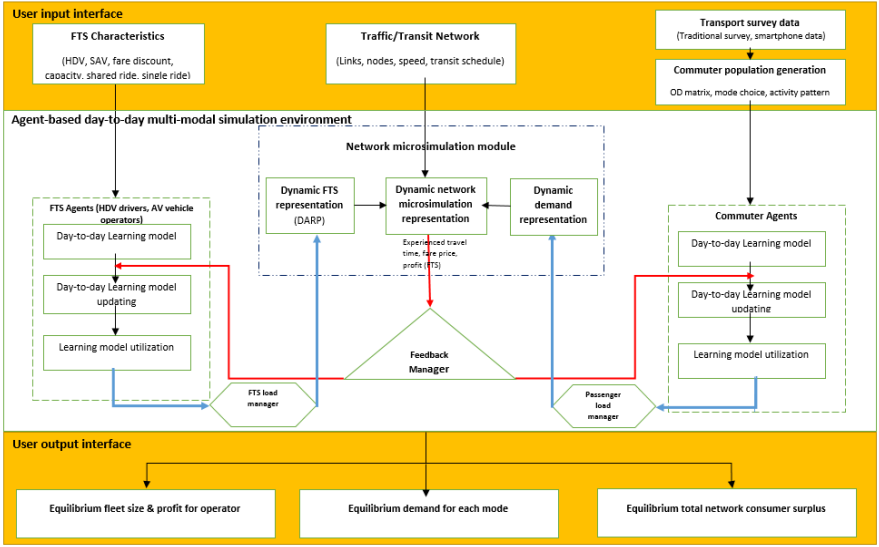

*Figure 10.1. Agent-based transportation simulation tool framework*

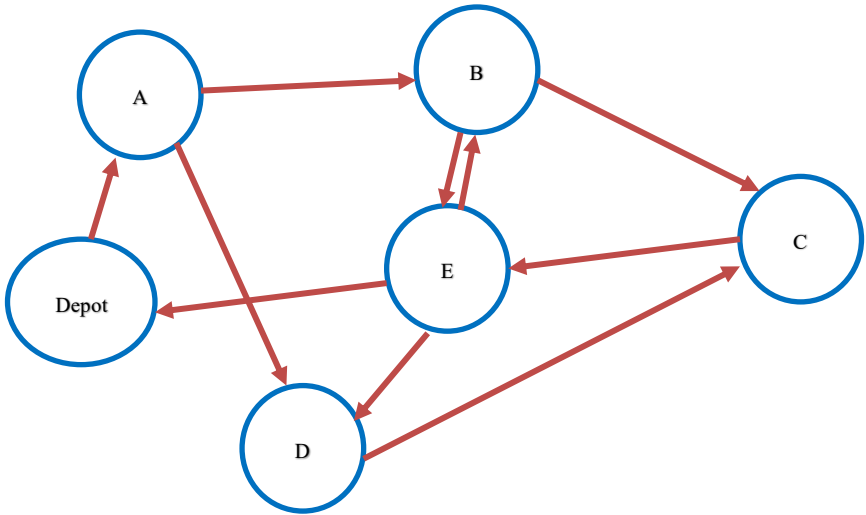

*Figure 10.2 Sample network*

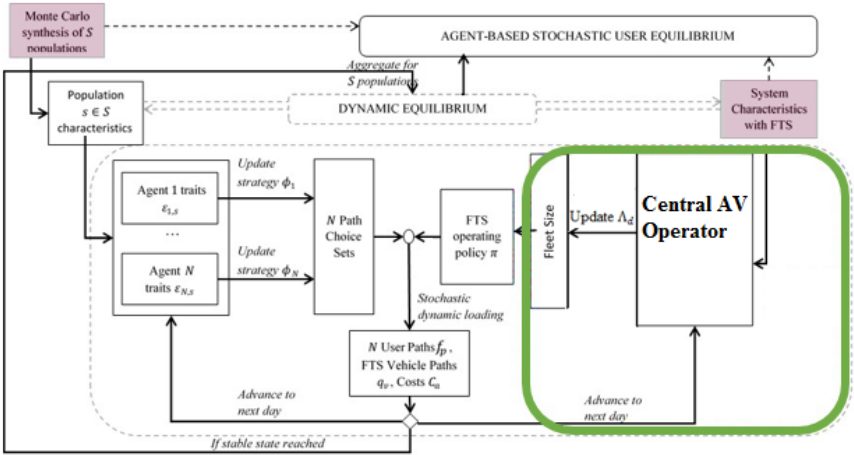

*Figure 10. 3 Agent-based day-to-day SUE framework for centrally operated AV fleets*

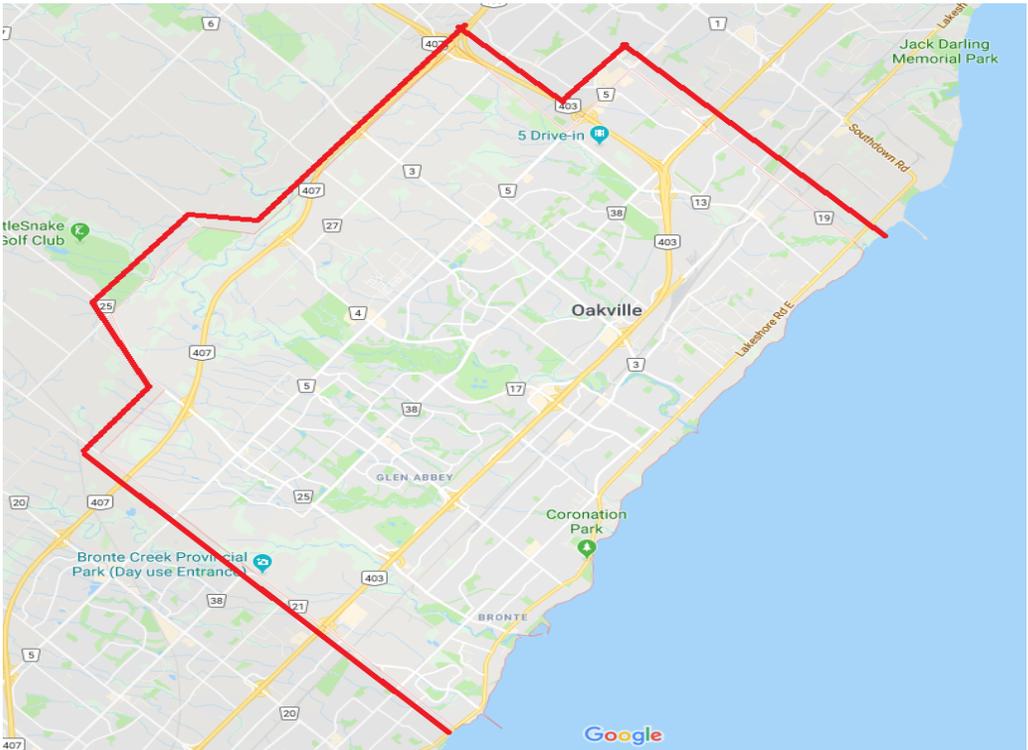

*Figure 10.4 Oakville Network*

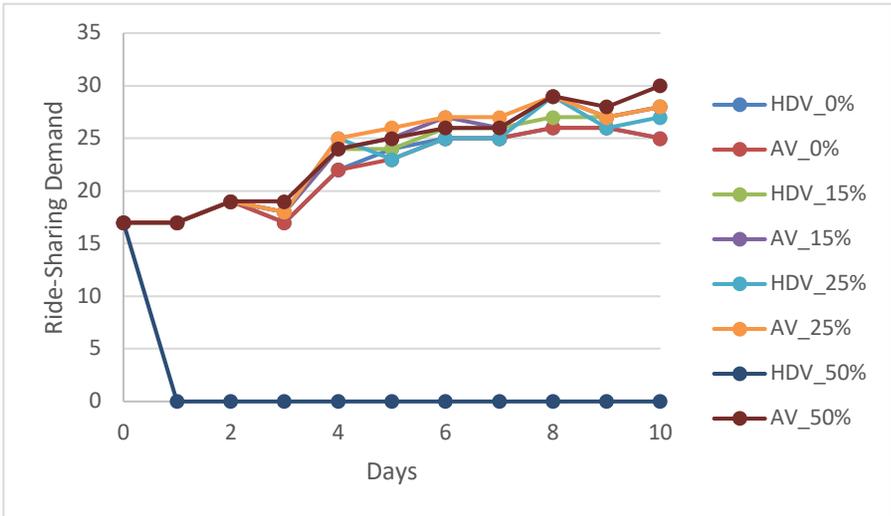

*Figure 10.5 Ride-sharing demand*

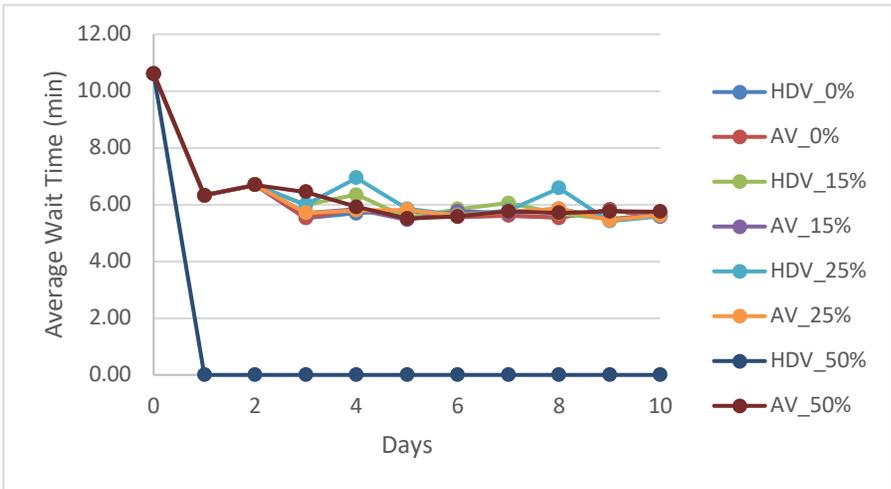

*Figure 10.6 Average wait time (min)*

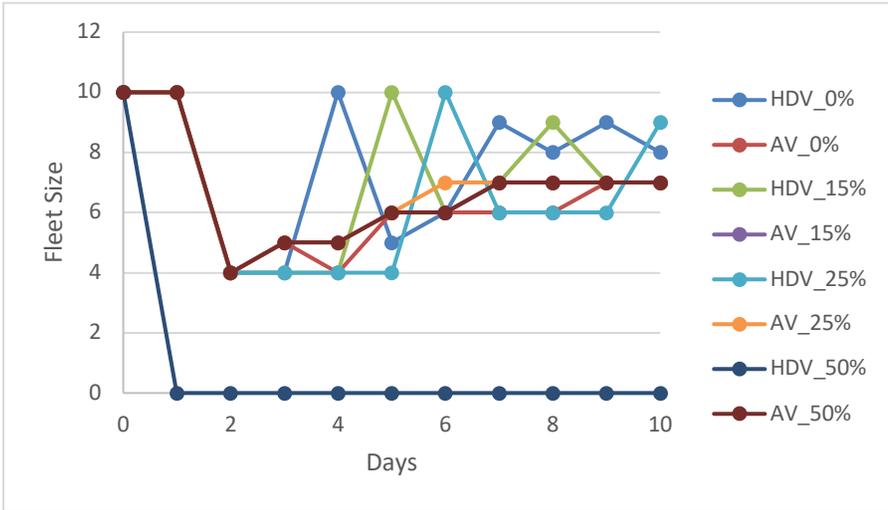

*Figure 10.7 Fleet size*

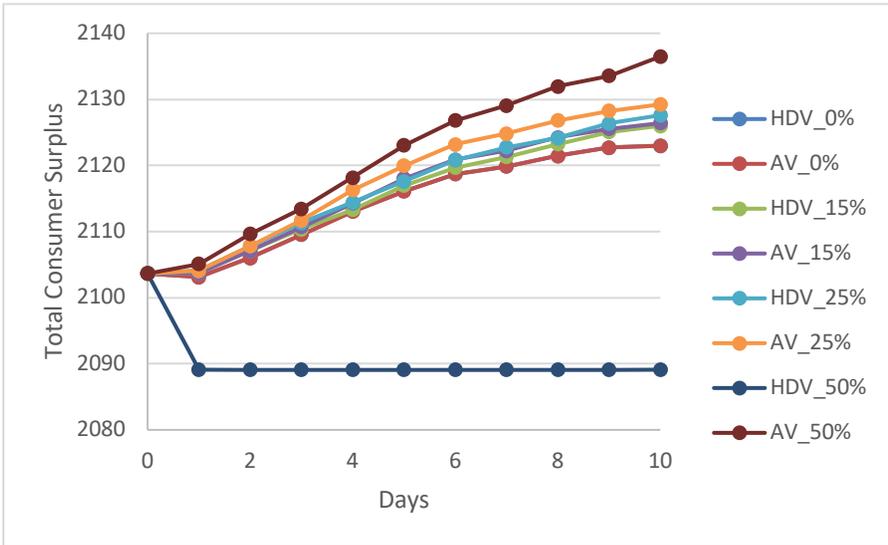

*Figure 10.8 Total consumer surplus*

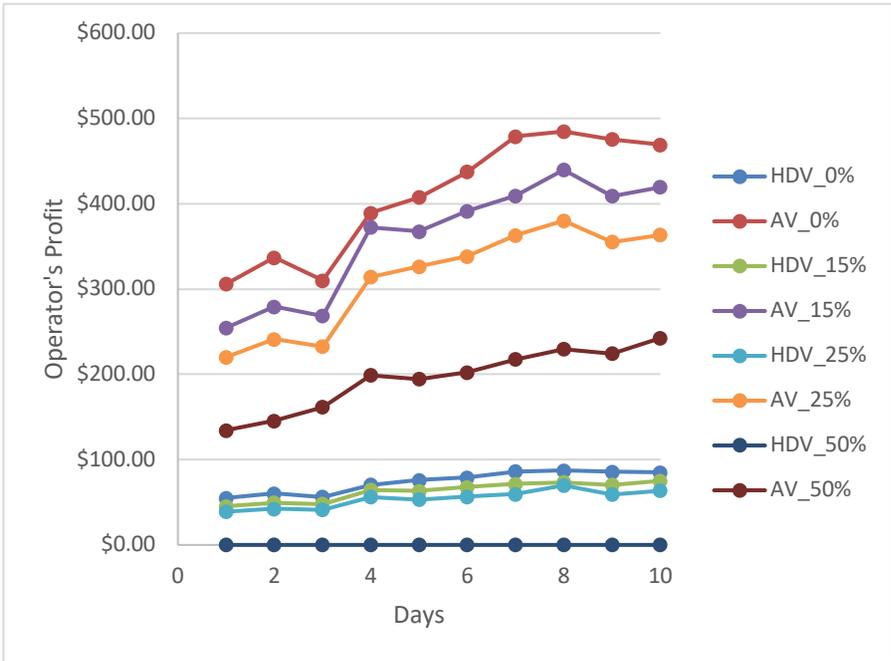

*Figure 10.9 Total Profit*